\documentclass{emulateapj}

\usepackage{amsmath}

\newcommand{\Mej}{$M_{\mathrm{ej}}$ }
\newcommand{\Ekin}{$E_{\mathrm{kin}}$ }
\newcommand{\Msun}{$M_{\odot}$ }
\newcommand{\Ni}{$^{56}$Ni }
\newcommand{\kms}{\mathrm{km\ s^{-1}}}
\def\ion#1#2{{\rm #1}~{\sc #2}}

\shorttitle{Fallback SNe: A Possible Origin of Peculiar SNe}
\shortauthors{Moriya et al.}

\begin{document}

\title{
Fallback Supernovae:
A Possible Origin of Peculiar Supernovae
with Extremely Low Explosion Energies
}

\def\ipmu{1}
\def\ut{2}
\def\resceu{3}
\def\konan{4}
\def\sth{5}
\def\mpa{6}
\def\pisa{7}
\def\inaf{8}

\author{
{Takashi Moriya}\altaffilmark{\ipmu,\ut,\resceu},
{Nozomu Tominaga}\altaffilmark{\konan,\ipmu},
{Masaomi Tanaka}\altaffilmark{\ipmu},
{Ken'ichi Nomoto}\altaffilmark{\ipmu,\ut},
{Daniel N. Sauer}\altaffilmark{\sth},\\
{Paolo A. Mazzali}\altaffilmark{\mpa,\pisa,\inaf},
{Keiichi Maeda}\altaffilmark{\ipmu}, and
{Tomoharu Suzuki}\altaffilmark{\ut}
}

\altaffiltext{\ipmu}{
Institute for the Physics and Mathematics of the Universe, University of Tokyo, Kashiwanoha 5-1-5, Kashiwa, Chiba 277-8583, Japan;
takashi.moriya@ipmu.jp
}
\altaffiltext{\ut}{
Department of Astronomy, Graduate School of Science, University of Tokyo,
Bunkyo-ku, Tokyo 113-0033, Japan
}
\altaffiltext{\resceu}{
Research Center for the Early Universe,
Graduate School of Science, University of Tokyo,
Bunkyo-ku, Tokyo 113-0033, Japan
}
\altaffiltext{\konan}{
Department of Physics, Faculty of Science and Engineering, Konan University, 8-9-1 Okamoto, Kobe, Hyogo 658-8501, Japan
}
\altaffiltext{\sth}{
Department of Astronomy, Stockholm University, Albanova University Center, SE-106 91 Stockholm, Sweden
}
\altaffiltext{\mpa}{
Max Planck Institute for Astrophysics,
Karl-Schwarzschild-Stra$\ss$e 1, 85741 Garching, Germany
}
\altaffiltext{\pisa}{
Scuola Normale Superiore, Piazza dei Cavalieri, 7, 56126 Pisa, Italy
}
\altaffiltext{\inaf}{
INAF-OAPd, vicolo dell'Osservatorio, 5, 35122 Padova, Italy
}

\begin{abstract}
We perform hydrodynamical calculations of core-collapse supernovae
(SNe) with low explosion energies.
These SNe do not have enough energy to eject the whole progenitor
and most of the progenitor falls back to the central remnant.
We show that such fallback SNe can have a variety of
light curves (LCs) but their photospheric velocities can only have
some limited values with lower limits.
We also perform calculations of nucleosynthesis and LCs
of several fallback SN model, and find that 
a fallback SN from the progenitor with 
a main-sequence mass of 13 \Msun can
account for the properties of the peculiar Type Ia supernova SN 2008ha.
The kinetic energy and ejecta mass of the model are
$1.2\times 10^{48}$ erg and 0.074 $M_\odot$, respectively, and
the ejected \Ni mass is 0.003 $M_\odot$. 
Thus, SN 2008ha can be a core-collapse SN with a large amount of
fallback.
We also suggest that SN 2008ha could have been accompanied with
long gamma-ray bursts and
long gamma-ray bursts without associated SNe
may be accompanied with very faint SNe with
significant amount of fallback which are similar to SN 2008ha.
\end{abstract}

\keywords{
gamma-ray burst: general -- supernovae: general --
supernovae: individual (SN 2008ha)
}

\section{Introduction}
A massive star with main-sequence mass above $\sim10$ \Msun
is thought to end its
life as a supernova (SN) after forming an Fe core at its center.
The SN is triggered by the gravitational collapse of the Fe core, thus being
called a core-collapse SN.
The mechanism that
leads to the final emergence of an SN from the collapse is still
under debate but observations show
that ejecta of the SN normally has a kinetic energy of $\sim~10^{51}$ erg.
Currently, however, theoretical attempts to simulate the whole explosion of a
core-collapse SN have not obtained
explosion energy as large as $10^{51}$ erg
(see, e.g., Janka et al. 2007; Bruenn et al. 2009;
Burrows et al. 2007; Suwa et al. 2009).

If the explosion energy is low,
the inner part of the 
star falls back onto the central remnant
and only the outer part of the star overcomes the gravitational potential.
The idea of the fallback was first introduced by Colgate (1971) and
many studies have since
investigated the effects of the fallback onto the central
remnant, e.g., black hole formation
({\it e.g.,} Chevalier 1989; Woosley \& Weaver 1995;
Fryer 1999; MacFadyen et al. 2001; Zhang et al. 2008).
Recently, more attention has been
paid to the outer part, which eventually
escapes from fallback, thus being ejected.
The ejecta might be observed as an SN ({\it e.g.,} Fryer et al. 2007, 2009)
and could produce the peculiar chemical abundance 
patterns of extremely metal-poor stars ({\it e.g.,} Iwamoto et al. 2005).
As this ejecta has a
kinetic energy just above the value required to overcome
the gravitational potential, it is expected to have very low energy.
If enough amount of $^{56}$Ni is also ejected or the ejecta interacts with the
circumstellar medium (Fryer et al. 2009), this ejecta might be
observed as an SN having very low line velocities. 

In this connection, the peculiar SN 2008ha 
is a suitable object, with sufficient observational data
that can be compared with the fallback SN models.
SN 2008ha was discovered on 2008 November UT 7.17
(Puckett, Moore, Newton, \& Orff 2008) and was found to be
one of the faintest SNe ever discovered (Valenti et al. 2009, hereafter V09;
Foley et al. 2009, hereafter F09).
It was found in an irregular galaxy UGC 12682
at a distance modulus of $\mu=31.64$ mag (F09).
Adopting a galactic extinction of $E(B-V)=0.08$ mag and little host
extinction, the peak absolute
$V$-band magnitude was found to be as faint as $-14.21 \pm 0.15$
mag (F09). From its spectral similarities to SN 2002cx-like Type Ia SNe,
SN 2008ha was classified as a peculiar Type Ia SN.

SN 2002cx-like SNe make a class of peculiar Type Ia SNe
(see, e.g., the supplementary information of V09;
SN 2002cx (Li et al. 2003; Jha et al. 2006) and SN 2005hk
(Phillips et al. 2007; Sahu et al. 2008)
are well-studied examples of this class).
Their spectra do not have strong absorptions of Si and S at early epochs
\footnote{However, the earliest observed spectrum of SN 2008ha before
the maximum luminosity showed these features (Foley et al. 2010),
although these features disappeared soon (V09; F09).},
which are the characteristic features of normal Type Ia SNe.
Line velocities of SN 2002cx-like SNe are very low
compared with normal Type Ia SNe (Branch et al. 2004).
They also have peculiar light curves (LCs), which decline slowly in spite
of their low maximum luminosities and do not show a
second peak which appears in the $I$ and $R$ band LCs of normal Type Ia SNe.

SN 2008ha has additional peculiarities.
The rise time of SN 2008ha
is faster than that of normal Type Ia SNe and the decline of the LC
after the maximum is very rapid: $\Delta m_{15}(B)=2.17\pm0.02$ mag
(F09)\footnote{$\Delta m_{15}(B)$ is 
the decline of the $B$ band magnitude in 15 days
since the $B$-band maximum.}.
Line velocities of SN 2008ha are as low as
$\sim 2,000\ \mathrm{km\ s^{-1}}$
around the maximum brightness (V09; F09).
Thus, the ejecta
is expected to have very low energy.
V09 suggested that the ejecta mass is \Mej = $0.1-0.5\ M_\odot$
and the kinetic energy is \Ekin = $(1-5)\times 10^{49}$ erg, while
F09 estimated \Mej= 0.15 \Msun and \Ekin= $2.3\times10^{48}$ erg.
The estimated mass of the ejected
\Ni is also as small as $(3-5)\times10^{-3}$ \Msun (V09) and
$(3.0\pm0.9)\times10^{-3}$ \Msun (F09).

Given the low energy and the small mass of the ejecta
as estimated from its spectral
features and LC shape as well as its star-forming host galaxy,
V09 concluded that
SN 2008ha is not a thermonuclear explosion but a core-collapse SN
with fallback.
F09 also pointed out the possibility of the core-collapse
origin but did not exclude the possibility of a thermonuclear explosion.
Indeed, based on the earliest spectrum observed, Foley et al. (2010)
suggested that SN 2008ha is related to a thermonuclear explosion.
Alternatively, Pumo et al. (2009) related SN 2008ha to an electron capture SN
(Nomoto 1984).

In this paper, we show that
the properties of SN 2008ha can be explained well by 
a fallback SN model.
We first perform numerical
calculations of hydrodynamics and nucleosynthesis
for several progenitor models.
Then, we perform radiative transfer calculations
to obtain the bolometric LCs and photospheric velocities
of these models to compare them with the observations of SN 2008ha.

In Section 2, we introduce the pre-SN models.
Methods used in our calculations
of hydrodynamics, nucleosynthesis, and bolometric LCs are
described in Section 3.
We show our results of hydrodynamical calculations in Section 4.
The results are compared
with the observed bolometric LC and photospheric velocity of
SN 2008ha in Section 5.
Discussion and conclusions are given in Section 6.

\section{Presupernova Models}
The spectra of SN 2008ha do not show hydrogen lines,
which implies the progenitor of SN 2008ha has lost
its H-rich envelope before the explosion.
In addition, the identification of
He lines in the spectra of SN 2008ha is very difficult
because the lines are overcrowded
(see the supplementary Figure 4 of V09 and Figure 8 of F09
for line identifications).
Thus, both a helium star and a carbon + oxygen (CO) 
star are possible
candidates for the progenitor of SN 2008ha.

We use pre-SN models of solar metallicity with
main-sequence masses
of 13 $M_\odot$, 25 $M_\odot$, and 40 \Msun
calculated by Umeda \& Nomoto (2002, 2005).
As these models have a H-rich envelope,
a He star is constructed by assuming that the whole H-rich envelope is lost
either by stellar wind or Roche-lobe overflow of a close binary,
and only
the He core remains at the pre-SN stage.
The boundary between the He core and the H-rich envelope is
assumed to be at the location $X$(H)$=0.1$ (hereafter,
 $X$(M) denotes the mass fraction of the element M).
We adopt the 25 \Msun and 40 \Msun models to construct the He star models
25He and 40He, respectively.
The He core masses
of 25He and 40He are $7.0$ \Msun and $15$ $M_\odot$, respectively.

\begin{figure}[t]
\begin{center}
\epsscale{1.2}
\plotone{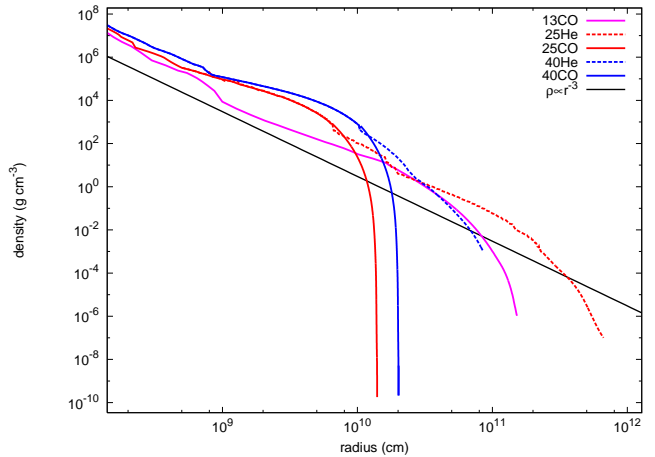}
\caption{
Density structure of the progenitor models.
The line of $\rho\propto r^{-3}$ is also shown for comparison.
}\label{DS}
\end{center}
\end{figure}

The CO star models are constructed by assuming that both the H-rich
and He envelopes are lost and
a CO core remains.
The boundary between the He envelope and the
CO core is set at $X$(He)$=0.1$.
We construct CO star models, 13CO, 25CO, and 40CO, from the
13 $M_\odot$, 25 $M_\odot$, and 40 \Msun models, respectively.
The CO core masses of 13CO, 25CO and 40CO are $2.7$ $M_\odot$, $5.7$ $M_\odot$,
and $14$ $M_\odot$, respectively. The density structures of all the models
used in this paper are shown in Figure \ref{DS}.

\section{Methods}
\subsection{Hydrodynamics and Nucleosynthesis}
Calculations of hydrodynamics and nucleosynthesis
are performed by using a spherical Lagrangian
hydrodynamic code with a piecewise parabolic method (Colella \& Woodward 1984).
The calculation of explosive
nucleosynthesis is coupled with hydrodynamics and the adopted
 reaction network includes 13 $\alpha$-particles,
i.e., $^{4}$He, $^{12}$C,
$^{16}$O, $^{20}$Ne, $^{24}$Mg, $^{28}$Si, $^{32}$S, $^{36}$Ar, $^{40}$Ca,
$^{44}$Ti, $^{48}$Cr, $^{52}$Fe, and $^{56}$Ni (see also Nakamura et al. 2001
for details).
The main purpose of the nucleosynthesis calculations is to see how much
$^{56}$Ni is produced at the explosion.
For this purpose, inclusion of only $\alpha$-nuclei is
a good approximation because 
$\alpha$-nuclei are the predominant yields of SNe.
The equation of state takes into account gas, radiation,
Coulomb interactions between ions
and electrons, $e^--e^+$ pair (Sugimoto \& Nomoto 1975),
and phase transition (Nomoto 1982; Nomoto \& Hashimoto 1988).
To obtain many $(E_\mathrm{kin}-M_\mathrm{ej})$ relations,
we compute several hydrodynamical models without
following nucleosynthesis and 
including only gas and radiation in the equation of state.
The omitted physics in the equation of state, such as Coulomb interaction,
mainly affects the result of nucleosynthesis and does not have
much effect on \Ekin and $M_\mathrm{ej}$.

As the explosion  mechanism of core-collapse SNe is
not yet clear, we initiate the
 explosion as a thermal bomb (e.g., Nakamura et al. 2001).
We put the thermal energy at $M_r=1.4\ M_\odot$
in the nucleosynthesis calculations, assuming that the 1.4 $M_\odot$
neutron star is initially formed and
the central remnant is treated as a point gravitational source.
Here, $M_r$ is the mass coordinate from the center.
There exist several ways to induce SN explosions,
e.g., a kinetic piston (e.g., Woosley \& Weaver 1995),
but it is suggested that the
results of nucleosynthesis are not sensitive to how energy is injected
(Aufderheide et al. 1991).
However, note that Young \& Fryer (2007) reports that the amount of fallback in
low energy explosions depends on the method by which explosions are indeuced.
Generally, explosions by kinetic piston have less fallback
because they tend to create stronger shocks
and thus, the difference in the method also affects the yields of
the nucleosynthesis.

\begin{figure}[t]
\begin{center}
\epsscale{1.2}
\plotone{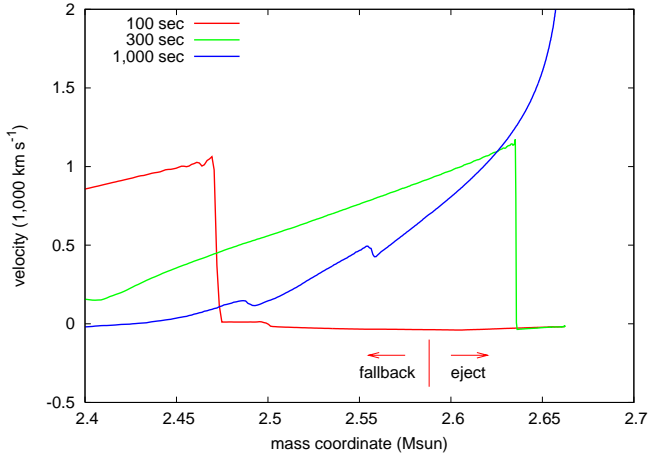}
\caption{
Change in the velocity profile shows the
shock propagation for the hydrodynamical model 13CO\_2.
The time since the explosion is shown.
The mass cut between the ejecta and the fallback material
is determined by whether the point exceeds the escape
 velocity or not.
}\label{shock}
\end{center}
\end{figure}

\subsection{Bolometric Light Curve}
Bolometric LCs are calculated by using an LTE radiative transfer code (Iwamoto
et al. 2000). This code assumes a gray atmosphere for
the $\gamma$-ray transport.
For the optical radiation transport,
electron scattering and line opacities are
taken into account.
Electron number density is evaluated by solving
the Saha equation. For simplicity,
the line opacity is assumed to be a constant
$0.06\ \mathrm{cm^2\ g^{-1}}$. This value has been
previously used for the explosion of CO stars (Maeda et al. 2003a).
The gray $\gamma$-ray opacity is set to be $0.027\ \mathrm{cm^2\ g^{-1}}$, which
is known to be a good approximation (Axelrod 1980).
Positrons emitted by the decay of $^{56}$Co are assumed to be
trapped in situ.
To compare with the computed bolometric LCs,
the observed bolometric LC of SN 2008ha is constructed as
shown in the Appendix.

\section{Results of Hydrodynamical Calculations}\label{hydro}
\subsection{$E_{\mathrm{kin}}-M_{\mathrm{ej}}$ Relation}
We calculate the hydrodynamics of 
the explosions and fallback for
25He, 40He, 13CO, 25CO, and 40CO with various
input energies.
In Figure \ref{shock}, the change in the velocity profile
shows the propagation of the shock wave.
When the explosion energy \Ekin is low,
the inner part of the progenitor cannot
overcome the gravitational potential
provided by the central remnant and thus falls back on to the remnant.
In such cases, only the outer layers of the progenitor are ejected.
The boundary between the fallback region and the ejecta is determined by whether
the velocity of the region exceeds the escape velocity.
We set a mass cut at this boundary to determine $M_\mathrm{ej}$.
The expansion of the region
above this mass cut eventually becomes
homologous. These homologous models are used for the calculations
of the bolometric LCs.

\begin{figure}[t]
\begin{center}
\epsscale{1.2}
\plotone{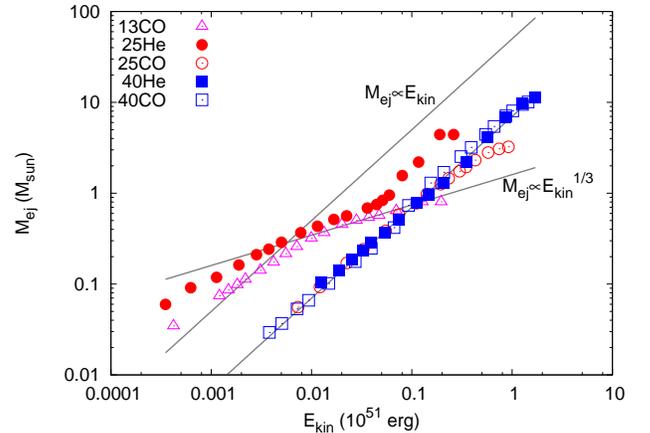}
\caption{Results of hydrodynamical calculations in the diagram of
the kinetic energy of the ejecta ($E_{\mathrm{kin}}$) and the
ejecta mass ($M_{\mathrm{ej}}$).
$E_{\mathrm{kin}}$ is the final kinetic energy at
infinity. For comparison, the lines of
$M_{\mathrm{ej}}\propto E_{\mathrm{kin}}$ and
$M_{\mathrm{ej}}\propto E_{\mathrm{kin}}^{1/3}$ are also shown.
}\label{EtoM}
\end{center}
\end{figure}

In Table \ref{tbl1}, we summarize \Ekin and \Mej for all our
hydrodynamical models and plot them in Figure \ref{EtoM}.
All the models with fallback are found to be on the line of either
$M_\mathrm{ej} \propto E_\mathrm{kin}$
or $M_\mathrm{ej}\propto E_\mathrm{kin}^{1/3}$.
The reason why there exist two relations between \Ekin and \Mej
can be understood as due to a difference in the density structure.
The density structure of the progenitor affects the manner of the shock
propagation, thus leading to the difference in the
$E_\mathrm{kin}-M_\mathrm{ej}$ relation.

Suppose the density structure of the progenitor is expressed as
$\rho\propto r^{-\alpha}$, where $\rho$ is the density and $r$ is the
radius. Sedov (1959) showed that a shock wave is accelerated when it is
propagating along the density structure with $\alpha > 3$ while it decelerates
when propagating along the density structure with $\alpha < 3$.
This means that to achieve a certain velocity at a place with a
density structure of $\alpha<3$, more energy is required than
the case of $\alpha>3$.
As the escape velocity determines the boundary between the ejecta and the
fallback region, the boundary is expected to be closer
to the central remnant
for the case for $\alpha>3$ than that of $\alpha<3$
if the same energy is injected.
Our hydrodynamical models show that if $\alpha > 3$ at the boundary between
the ejecta and the fallback region, the results follow the relation
$M_\mathrm{ej}\propto E_\mathrm{kin}$, and if $\alpha <3$, they follow
$M_\mathrm{ej}\propto E_\mathrm{kin}^{1/3}$.
The exact physical reason why \Mej and \Ekin are related as
$M_\mathrm{ej}\propto E_\mathrm{kin}$ and
$M_\mathrm{ej}\propto E_\mathrm{kin}^{1/3}$
is still unclear. We just treat it as an empirical relation
in this paper and leave it as an open question.

\begin{figure*}[htbp]
\begin{center}
  \begin{minipage}[t]{0.45\textwidth}
    \epsscale{1.2}
    \plotone{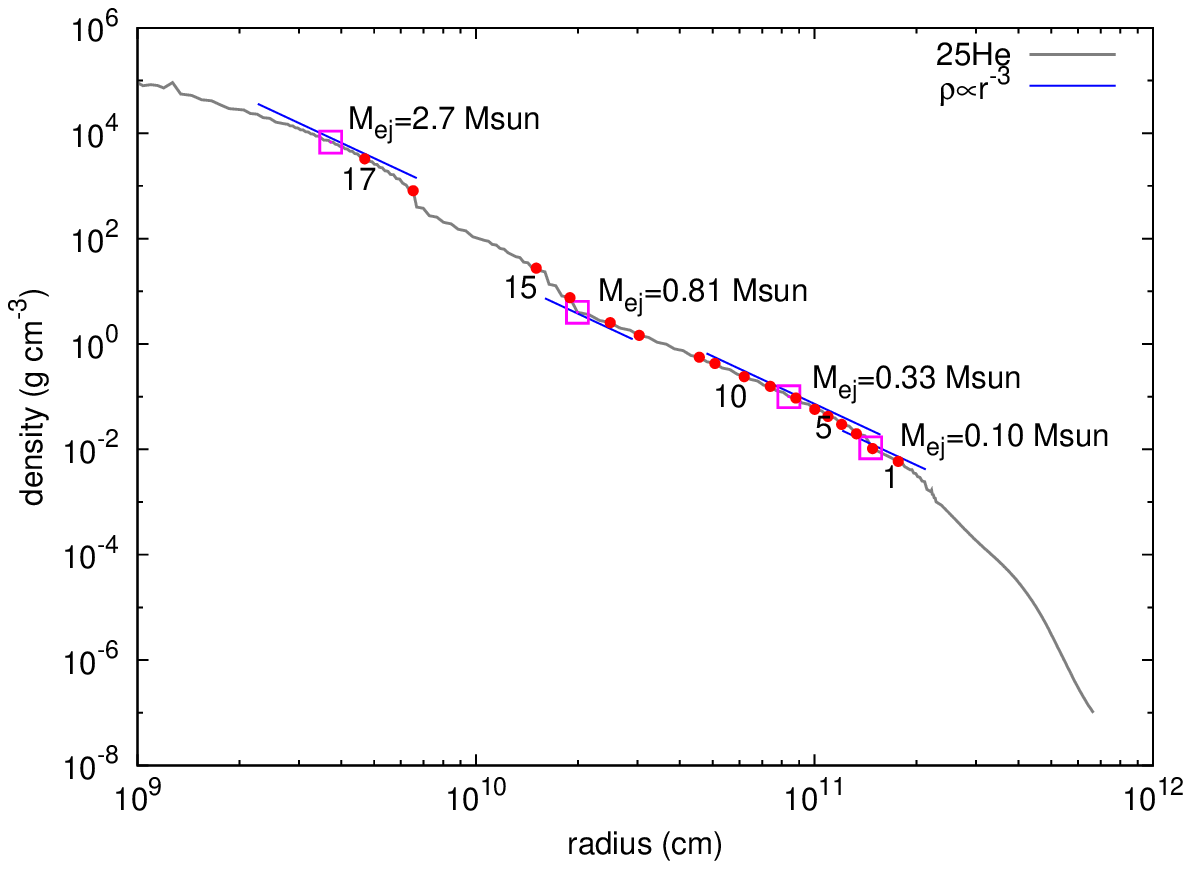}
  \end{minipage}
  \begin{minipage}[t]{0.45\textwidth}
    \epsscale{1.2}
    \plotone{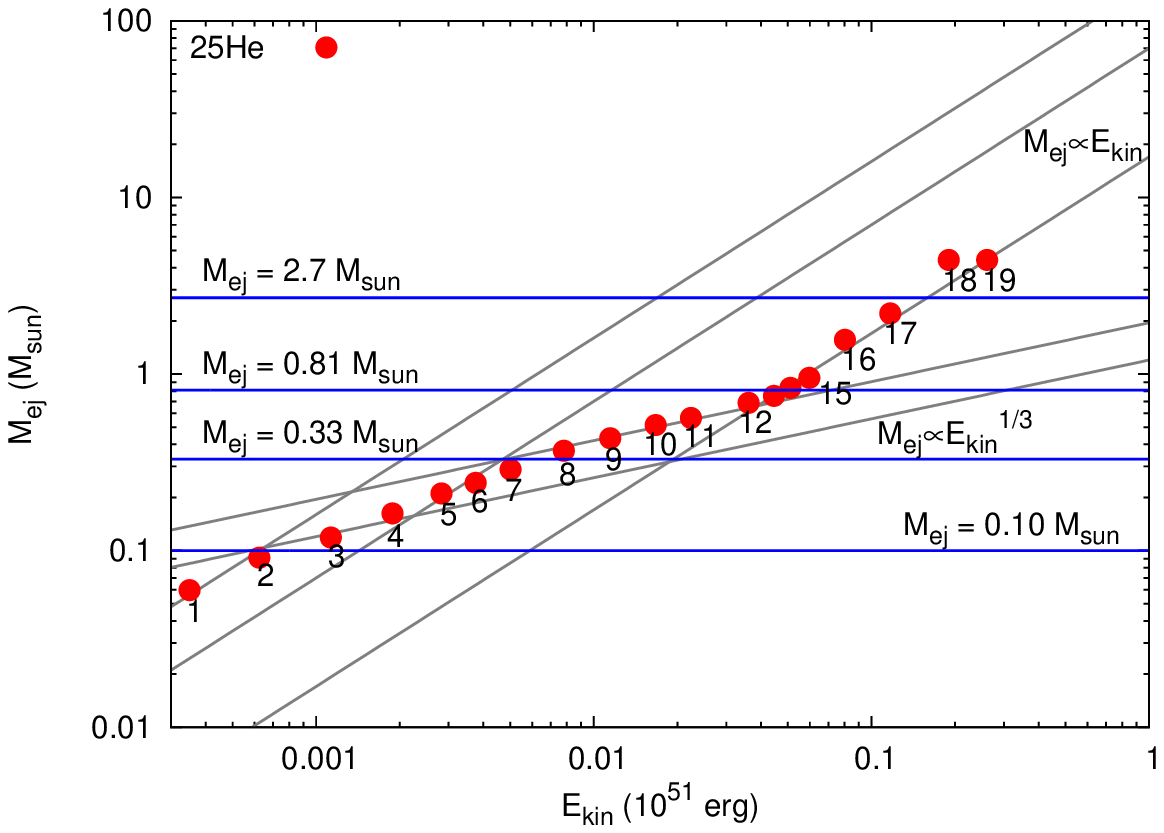}
  \end{minipage}
\caption{Density structure of 25He and
the $E_\mathrm{kin}-M_\mathrm{ej}$ relation of 25He.
Along the structure line, we plot the points
where the density slope is $\rho\propto r^{-3}$ (open squares).
In the right panel, we plot hydrodynamical models
(Nos. 1-19) for 25He listed in Table \ref{tbl1}.
The numbers attached to the points in the left panel
are the same model numbers as in the right panel. The location of the
model points indicates the mass cut of the hydrodynamical model.
Looking at the density structure from outside, 
there is a small region where
 the density structure follows $\alpha<3$
 which corresponds to model 3 with
$M_\mathrm{ej}\simeq 0.10\ M_\odot$. Thus,
the $E_\mathrm{kin}-M_\mathrm{ej}$ relation of the models around this
region follows
$M_\mathrm{ej}\propto E_\mathrm{kin}^{1/3}$. At
 $M_\mathrm{ej}\simeq 0.33-0.81\ M_\odot$,
the corresponding models 8-13 have a density structure
with $\alpha<3$, thus following $M_\mathrm{ej}\propto E_\mathrm{kin}^{1/3}$.
At larger $M_\mathrm{ej}$, models 14-17 have a density 
structure with $\alpha>3$ and follow
$M_\mathrm{ej}\propto E_\mathrm{kin}$.
The last two models, 18 and 19, have no fallback.
}
\label{25He}
\end{center}
\end{figure*}

\begin{deluxetable*}{ccccccccccc}
\tablecaption{Results of Hydrodynamical Calculations}
\tablehead{
&
\multicolumn{2}{c}{13CO}&
\multicolumn{2}{c}{25He}&
\multicolumn{2}{c}{25CO}&
\multicolumn{2}{c}{40He}&
\multicolumn{2}{c}{40CO}\\
No.&\Ekin&\Mej&\Ekin&\Mej&\Ekin&\Mej&\Ekin&\Mej&\Ekin&\Mej}
\startdata
1&$0.00042$&$0.035$&$0.00035$&$0.060$&
$0.0073$&$0.056$&$0.012$&$0.10$&$0.0038$&$0.029$\\
2&$0.0012$&$0.074$&$0.00062$&$0.091$&
$0.012$&$0.093$&$0.019$&$0.14$&$0.0050$&$0.037$\\
3&$0.0015$&$0.086$&$0.0011$&$0.12$&
$0.022$&$0.17$&$0.026$&$0.19$&$0.0072$&$0.053$\\
4&$0.0018$&$0.098$&$0.0019$&$0.16$&
$0.033$&$0.24$&$0.033$&$0.23$&$0.0094$&$0.066$\\
5&$0.0022$&$0.11$&$0.0028$&$0.21$&
$0.055$&$0.39$&$0.039$&$0.29$&$0.015$&$0.10$\\
6&$0.0031$&$0.14$&$0.0038$&$0.24$&
$0.074$&$0.58$&$0.054$&$0.37$&$0.027$&$0.18$\\
7&$0.0042$&$0.17$&$0.0050$&$0.29$&
$0.11$&$0.78$&$0.073$&$0.51$&$0.039$&$0.25$\\
8&$0.0055$&$0.22$&$0.0078$&$0.37$&
$0.14$&$0.98$&$0.11$&$0.78$&$0.066$&$0.42$\\
9&$0.0071$&$0.26$&$0.011$&$0.43$&
$0.19$&$1.3$&$0.15$&$0.95$&$0.093$&$0.73$\\
10&$0.010$&$0.32$&$0.017$&$0.51$&
$0.23$&$1.5$&$0.21$&$1.2$&$0.15$&$1.3$\\
11&$0.013$&$0.37$&$0.022$&$0.56$&
$0.30$&$1.7$&$0.35$&$2.2$&$0.21$&$1.7$\\
12&$0.020$&$0.45$&$0.036$&$0.69$&
$0.35$&$1.9$&$0.56$&$4.1$&$0.31$&$2.5$\\
13&$0.028$&$0.50$&$0.045$&$0.75$&
$0.43$&$2.7$&$0.86$&$6.8$&$0.39$&$3.2$\\
14&$0.037$&$0.54$&$0.051$&$0.83$&
$0.58$&$2.8$&$1.3$&$9.7$&$0.54$&$4.4$\\
15&$0.047$&$0.57$&$0.060$&$0.95$&
$0.74$&$3.1$&$1.7$&$11$&$0.66$&$5.4$\\
16&$0.070$&$0.64$&$0.080$&$1.6$&
$0.92$&$3.2$&$\ldots$&$\ldots$&$0.87$&$7.1$\\
17&$\ldots$&$\ldots$&$0.12$&$2.2$&
$\ldots$&$\ldots$&$\ldots$&$\ldots$&$1.0$&$8.1$\\
18&$\ldots$&$\ldots$&$0.19$&$4.4^\mathrm{a}$&
$\ldots$&$\ldots$&$\ldots$&$\ldots$&$1.3$&$9.5$\\
19&$\ldots$&$\ldots$&$0.26$&$4.4^\mathrm{a}$&
$\ldots$&$\ldots$&$\ldots$&$\ldots$&$1.4$&$10$
\enddata
\tablenotetext{a}{No fallback}
\tablecomments{The units of \Ekin and \Mej are $10^{51}$ erg and
 $M_\odot$, respectively.}
\label{tbl1}
\end{deluxetable*}

\begin{figure}[t]
\begin{center}
\epsscale{1.2}
\rotatebox{0}{
\plotone{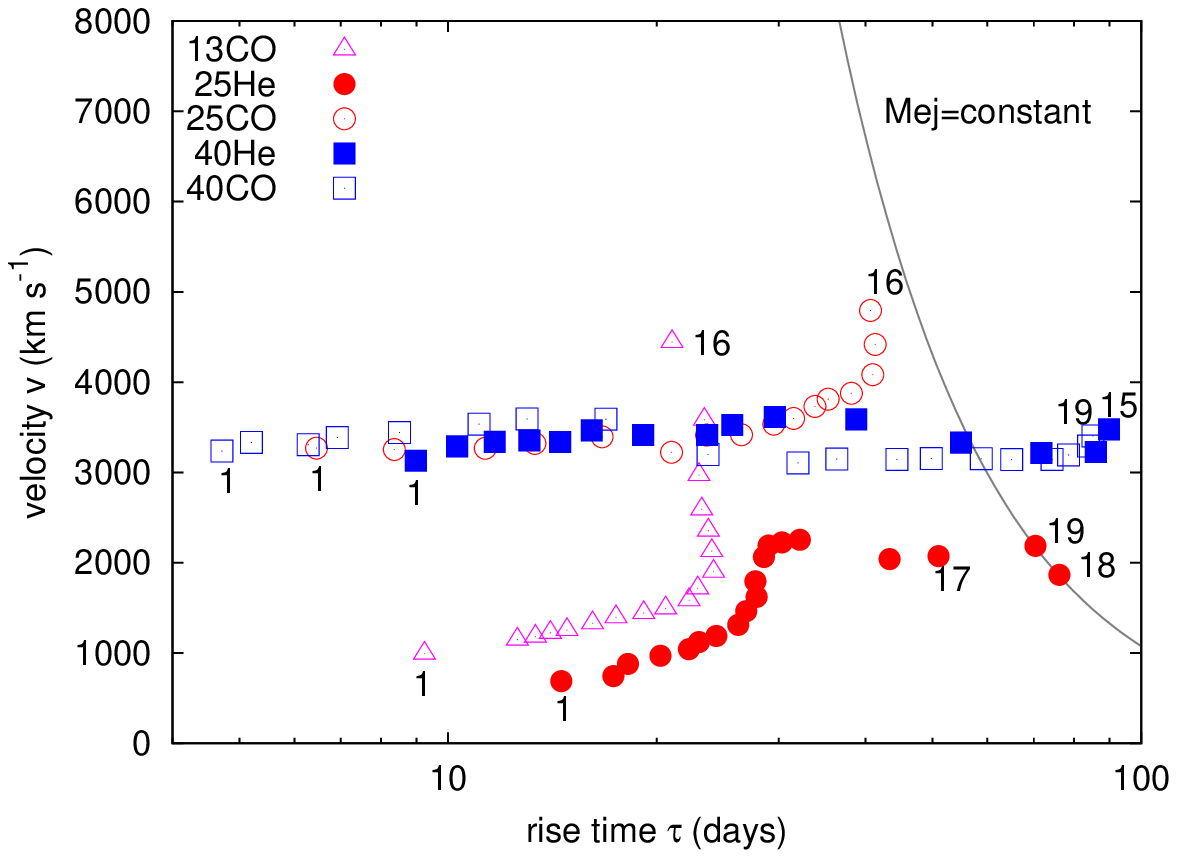}}
\caption{
Results of hydrodynamical calculations
in the diagram of the rise time ($\tau$) and the ejecta velocity ($v$)
defined by Equations (\ref{tau}) and (\ref{v}).
The attached numbers correspond to the numbers of the models
shown in Table \ref{tbl1}.
The line on the right-hand side is the line $M_\mathrm{ej}=\mathrm{constant.}$
}
\label{TtoV}
\end{center}
\end{figure}

Figure \ref{25He} (left) shows the density structure of the progenitor
model 25He. 
In Figure \ref{25He} (right), we plot hydrodynamical models
(Nos. 1-19) for 25He listed in Table \ref{tbl1}.
The numbers attached to the points in the left panel of the figure
are the same model numbers as in the right panel. The location of the
model points indicates the mass cut of the hydrodynamical model.
Looking at the density structure from outside, 
there is a small region where
 the density structure follows $\alpha<3$
 which corresponds to model 3 with
$M_\mathrm{ej}\simeq 0.10\ M_\odot$. Thus,
the $E_\mathrm{kin}-M_\mathrm{ej}$ relation of the models around this
region follows
$M_\mathrm{ej}\propto E_\mathrm{kin}^{1/3}$. At
 $M_\mathrm{ej}\simeq 0.33-0.81\ M_\odot$,
the corresponding models 8-13 have a density structure
with $\alpha<3$, thus following $M_\mathrm{ej}\propto E_\mathrm{kin}^{1/3}$.
At larger $M_\mathrm{ej}$, models 14-17 have a density 
structure with $\alpha>3$ and follow
$M_\mathrm{ej}\propto E_\mathrm{kin}$.
The last two models, 18 and 19, have no fallback.

For the extremely low $E_\mathrm{kin}$, the mass cut in
the explosion models lies in the outermost layer
where the density declines exponentially in all
the progenitor models.
Thus, the model sequences follow the relation
$E_\mathrm{kin}\propto M_\mathrm{ej}$ at the low \Ekin limit
in the $E_\mathrm{kin}-M_\mathrm{ej}$ plane.
This proportionality between \Ekin and \Mej was also shown by
Nadezhin \& Frank-Kamenetskii (1963)
in the context of nova explosions.
Thus, the ejecta velocity, which is scaled as
$v\propto \left(E_\mathrm{kin}/{M_\mathrm{ej}}\right)^{1/2}$, would 
be a constant, being independent of \Ekin for each progenitor model.
This means that however low $E_\mathrm{kin}$ is,
the ejecta velocity, i.e., the line velocities of the spectra, does
not become lower than a certain asymptotic value.

\subsection{Rise Time - Ejecta Velocity Relation}
Based on the $E_{\mathrm{kin}}-M_{\mathrm{ej}}$ relation, we can construct a relation between
observable quantities: the rise time ($\tau$) of the LC versus velocity ($v$)
of the ejecta.
The ejecta velocity approximates the line velocities
in the observed spectra.
For each set of ($E_{\mathrm{kin}}$, $M_{\mathrm{ej}}$), $\tau$ is derived from the relation
$\tau\propto \kappa^{1/2}(M_{\mathrm{ej}}^3/E_{\mathrm{kin}})^{1/4}$ (Arnett 1982), where
$\kappa$ is the total opacity and $v$ is simply scaled as
$v\propto(E_{\mathrm{kin}}/M_{\mathrm{ej}})^{1/2}$. 
For simplicity, we
assume that $\kappa$ is
constant for all the models.
For illustration, we choose the proportional constants in
$\tau$ and $v$ to match the typical values of Type Ia SNe,
$E_{\mathrm{kin}}=1.4\times10^{51}$
erg, $M_{\mathrm{ej}}=1.4\ M_\odot$, $\tau=19.5$ days, and $v=9000$ $\mathrm{km\ s^{-1}}$,
and get the relations:
\begin{eqnarray}
\tau&=&16.5\left(\frac{\left(M_{\mathrm{ej}}/M_\odot\right)^3}{E_{\mathrm{kin}}/10^{51}\mathrm{erg}}\right)^{1/4}\ \mathrm{days},
\label{tau}\\
v&=&9000\left(\frac{E_{\mathrm{kin}}/10^{51}\mathrm{erg}}{M_{\mathrm{ej}}/M_\odot}\right)^{1/2}\
 \mathrm{km\ s^{-1}}. \label{v}
\end{eqnarray}

In Figure \ref{TtoV} we plot $(\tau, v)$ for the model sequences in
Table \ref{tbl1} and Figure \ref{EtoM}.
As derived from Equations (\ref{tau}) and (\ref{v}), 
the models with $M_\mathrm{ej}\propto E_\mathrm{kin}$
are located along the line with
$\tau\propto M_\mathrm{ej}^{1/2}\propto E_\mathrm{kin}^{1/2}$
and $v\simeq\mathrm{constant}$,
while the models with $M_\mathrm{ej}\propto E_\mathrm{kin}^{1/3}$
are located along the line with
 $\tau\simeq\mathrm{constant}$
and $v\propto M_\mathrm{ej}\propto E_\mathrm{kin}^{1/3}$.
For large $E_\mathrm{kin}$,
all materials outside the proto-neutron star are ejected without
fallback
and $M_\mathrm{ej}=M_\mathrm{pro}-M_\mathrm{rem}$
is a constant, where $M_\mathrm{pro}$ is the progenitor's pre-SN mass and
$M_\mathrm{rem}$ is the mass of the central remnant below the mass cut.
As $M_\mathrm{ej}$ is a constant,
$\tau$ and $v$ follow the curve
$\tau^2v\propto M_{\mathrm{ej}}=\mathrm{constant}$,
as derived from Equations (\ref{tau}) and (\ref{v}).
The black curve on the right side of Figure \ref{TtoV}
shows the curve of $M_\mathrm{ej}=\mathrm{constant}$ for model 25He.

Our hydrodynamical models have a wide range of $\tau$
as seen in Figure \ref{TtoV} and it implies that
fallback SNe have a variety of LCs.
However, $v$ has only limited values for each progenitor
and the progenitor of a fallback SN can be constrained not by
its LC but by its photospheric velocity.
In the next section,
we constrain the progenitor of SN 2008ha mainly
by using its photospheric velocity.
As the observed line velocities around the maximum luminosity of SN 2008ha
are $\sim$ 2000 $\kms$, it is expected that 25CO, 40He, and 40CO
would have too high photospheric velocities ($v$) to
be consistent with those of SN 2008ha.\\

\section{SN 2008ha}
\begin{figure}[t]
\begin{center}
\epsscale{1.2}
\rotatebox{0}{
\plotone{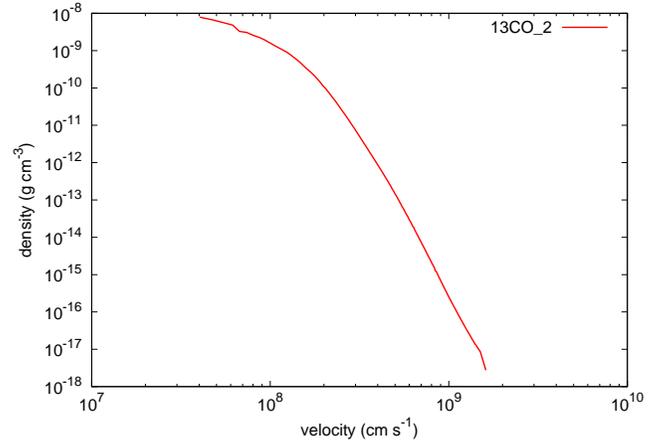}}
\caption{
Density structures of 13CO\_2 at 2 days after the explosion.
}\label{Gdens}
\end{center}
\end{figure}

\begin{figure}[t]
\begin{center}
\epsscale{1.2}
\rotatebox{0}{
\plotone{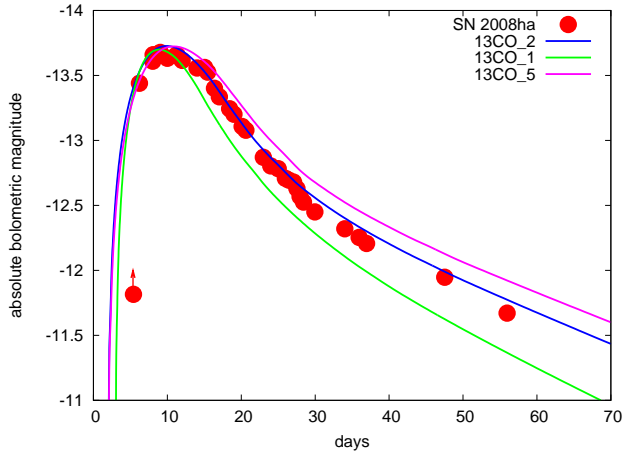}}
\caption{
Bolometric LC of SN 2008ha and the calculated bolometric LCs for some
 explosion models of 13CO are shown. The bolometric LC of SN 2008ha is constructed as
explained in the Appendix.
Horizontal axis represents days since the explosion of the theoretical
 LCs of 13CO\_2 and 13CO\_5. The explosion date of 13CO\_1 is
 + 1 day in the figure.
The kinetic energy and mass of ejecta of 13CO\_2 are 
\Ekin = $1.2\times10^{48}$ erg and \Mej = 0.074 $M_\odot$. The mass of
 \Ni is 0.003 $M_\odot$ in 13CO\_2. The rise time of 13CO\_2 is 9.8 days.
$E_\mathrm{kin}$ and $M_\mathrm{ej}$ of 13CO\_1 and 13CO\_5 are
$(E_\mathrm{kin},~M_\mathrm{ej})=(4.2\times10^{47}\mathrm{erg},0.035M_\odot)$ 
and $(2.2\times10^{48}\mathrm{erg},0.11M_\odot)$, respectively.
}\label{COLC}
\end{center}
\end{figure}

We calculate bolometric LCs and the photospheric
velocities for the models shown in Figure \ref{EtoM}.
We adopt the structure of the ejecta when it reaches the homologous expansion
in hydrodynamical calculations and assume that
$^{56}$Ni is uniformly mixed throughout the ejecta.
Among all the models shown in Figure \ref{EtoM},
we find that model 13CO\_2 with
$(E_\mathrm{kin}, M_\mathrm{ej})=(1.2\times10^{48}\ \mathrm{erg}, 0.074\ M_\odot)$
is consistent with both the
bolometric LC and the photospheric velocity of SN 2008ha.
Fallback occurs in 13CO\_2 and only the outermost layer of the
progenitor is ejected. The density structure of the model is shown
in Figure \ref{Gdens}.
Figure \ref{COLC} shows that the calculated bolometric LC of 13CO\_2
is in good agreement with the
observed bolometric LC of SN 2008ha (see the Appendix).
The \Ni mass ejected is assumed to be 0.003 $M_\odot$.
The rise time of 13CO\_2 is 9.8 days.
LCs from other explosion models of 13CO are also shown for comparison
in Figure \ref{COLC}.

\begin{deluxetable*}{ccccccccccccc}
\tablecaption{Uniform Composition Based on the Result of the Nucleosynthesis of 13CO\_2}
\tablehead{
$^{4}$He&$^{12}$C&$^{16}$O&$^{20}$Ne&$^{24}$Mg&$^{28}$Si&$^{32}$S&$^{36}$Ar&$^{40}$Ca&$^{44}$Ti&$^{48}$Cr&$^{52}$Fe&\Ni
}
\startdata
$0.037$&$0.20$&$0.32$&$0.14$&$0.066$&
$0.063$&$0.034$&$0.0065$&
$0.0045$&$0.00051$&$0.0017$&
$0.011$&$0.12$
\enddata
\tablecomments{The mass fraction of each element is shown.}
\label{tbl2}
\end{deluxetable*}

In the  nucleosynthesis calculation,
the explosion of 13CO\_2 produces 0.15 $M_\odot$ of $^{56}$Ni
at $M_r>1.4~M_\odot$.
If we assume uniform mixing of \Ni at $M_r>1.4\ M_\odot$,
the ejecta will contain 0.0086 $M_\odot$ $^{56}$Ni (Table \ref{tbl2}).
To reproduce the luminosity of SN 2008ha,
0.003 \Msun of \Ni needs to be contained in the ejecta.
This implies that mixing due to the Rayleigh-Taylor instability
(e.g., Hachisu et al. 1991; Joggerst et al. 2009)
or a jet (e.g., Maeda \& Nomoto 2003b; Tominaga 2009) occurs in SN 2008ha 
to bring \Ni to the outermost layer before the fallback.

\begin{figure}[t]
\begin{center}
 \epsscale{1.2}
 \plotone{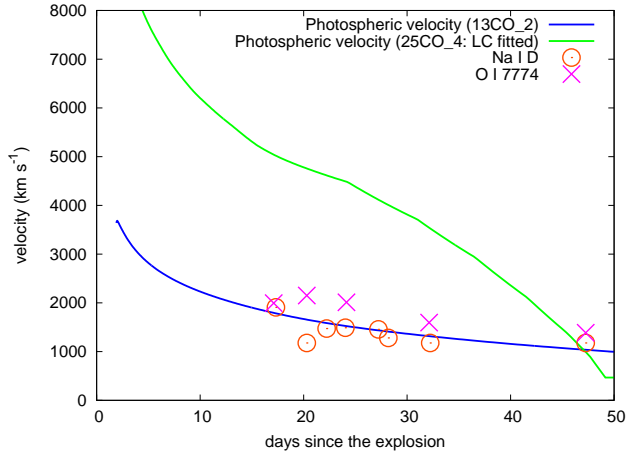}
\caption{
Photospheric velocities of 13CO\_2
obtained from the bolometric LC calculations.
These are compared with the observed line velocities
of \ion{Na}{i} D and \ion{O}{i} 7774 from Figure 5 of F09,
which are expected to be good tracers of the photosphere.
Days are measured since the explosion date of each model.
The rise time of 13CO\_2 is 9.8 days.
We also show the explosion model of 25CO which has \Ekin= $3.3\times 10^{49}$
 erg and \Mej= $0.24\ M_\odot$.
This model reproduces well the LC of SN 2008ha but
the photospheric velocity is too high to be comparable to the line
 velocities of SN 2008ha.
}\label{photo}
\end{center}
\end{figure}

Figure \ref{photo} shows the photospheric velocities of 13CO\_2 compared
with the observed line velocities of SN 2008ha (Figure 5 of F09).
Among the line velocities shown in F09,
we take \ion{Na}{i} D and \ion{O}{i} 7774
as good
tracers of the photospheric velocity because these lines can be clearly
distinguished and their line velocities are slower than other lines.
The evolution of the photospheric velocity of 13CO\_2 follows the line
velocities of these tracers well.
Thus, it is expected that our models will also be consistent with the
observed spectra of SN 2008ha. Detailed synthetic spectra based on
our model will be shown in a forthcoming paper (D. N. Sauer et al. 2010,
in preparation).

The model from 25CO whose photospheric velocity is shown in Figure
\ref{photo} has
\Ekin = $3.3\times10^{49}$ erg and \Mej = 0.24 $M_\odot$ (25CO\_4).
This model reproduce well the bolometric LC of SN 2008ha.
However, the photospheric velocity of this model
is much higher than the line velocities of SN 2008ha.
This result is expected from Figure \ref{TtoV} because
the ejecta velocities ($v$) of the explosion models of 25CO are too high
for SN 2008ha (Figure \ref{TtoV}).
On the other hand,
the explosion model of 25He with the smallest energy
we calculated ($E_\mathrm{kin}=3.5\times10^{47}$ erg) 
still has a long rise time compared to SN 2008ha.
Its LC is close to that of the explosion model 13CO\_5
shown in Figure \ref{COLC}, which has almost the same $\tau$
(Figure \ref{TtoV}).
It is expected from Figures \ref{EtoM} and \ref{TtoV} that
an explosion model of 25He 
with smaller energy 
could be consistent with SN 2008ha but the energy will
have to be quite small ($E_\mathrm{kin}\simeq10^{47}$ erg).
Still, there remains a possibility that such an SN with a very
small explosion energy could emerge as a result of the fallback.

Although we assume a large amount of mass loss during
the evolution of the progenitors, we do not assume the
existence of a circumstellar matters due to a mass loss
in our LC calculations.
If the circumstellar matters are dense enough,
SNe with fallback will be brightened by the
interaction between the ejecta of SNe and the circumstellar
matters (Fryer et al. 2009).
According to Fryer et al. (2009),
this interaction could result in SNe having
LCs with rising times and luminosities
similar to those of SN 2008ha.
Considering the fact that
 we do not see any evidence for the interaction
in the spectra of SN 2008ha and there is no
sudden drop in the tail of the LC of SN 2008ha, which is expected in
the interaction-powered SN models
and is naturally explained by
the nuclear decay of $^{56}$Co,
we think that SN 2008ha is likely to be powered by
the nuclear decay of \Ni
and the circumstellar matter around the progenitor
of SN 2008ha is so thin that the interaction does
not become the major energy source of the LC.
However, in conditions such as the mixing of the whole
exploding star does not occur,
almost all \Ni would fall
back to the central remnant
and only the interaction between the ejecta and the circumstellar
matter would be the energy source to brighten the SN.
We still do not have a clear observational SN with fallback
brightened by the interaction, but
such SNe might be discovered in the future.

\begin{figure*}[t]
\begin{center}
  \begin{minipage}[t]{0.45\textwidth}
    \epsscale{1.2}
    \plotone{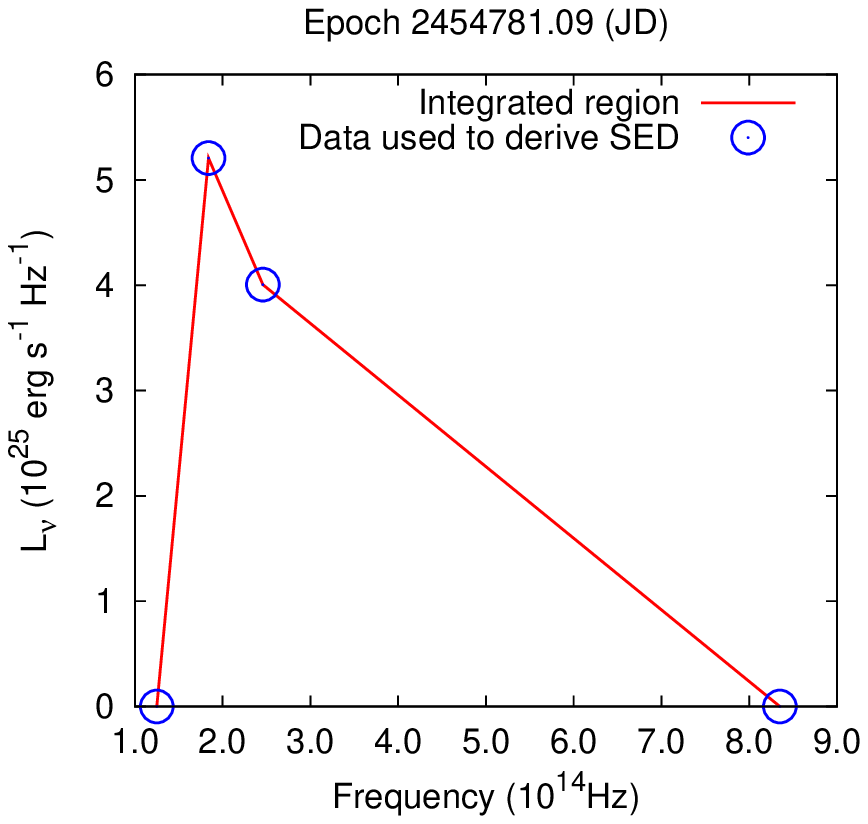}
	\end{minipage}
  \begin{minipage}[t]{0.45\textwidth}
    \epsscale{1.2}
    \plotone{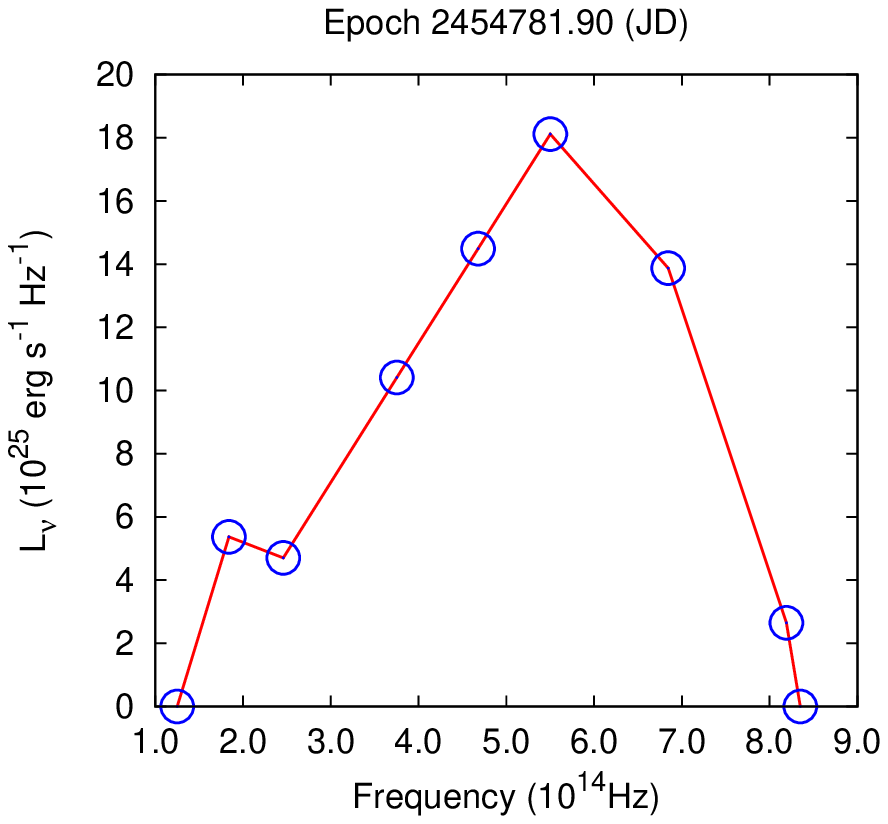}
	\end{minipage}
  \begin{minipage}[t]{0.45\textwidth}
    \epsscale{1.2}
    \plotone{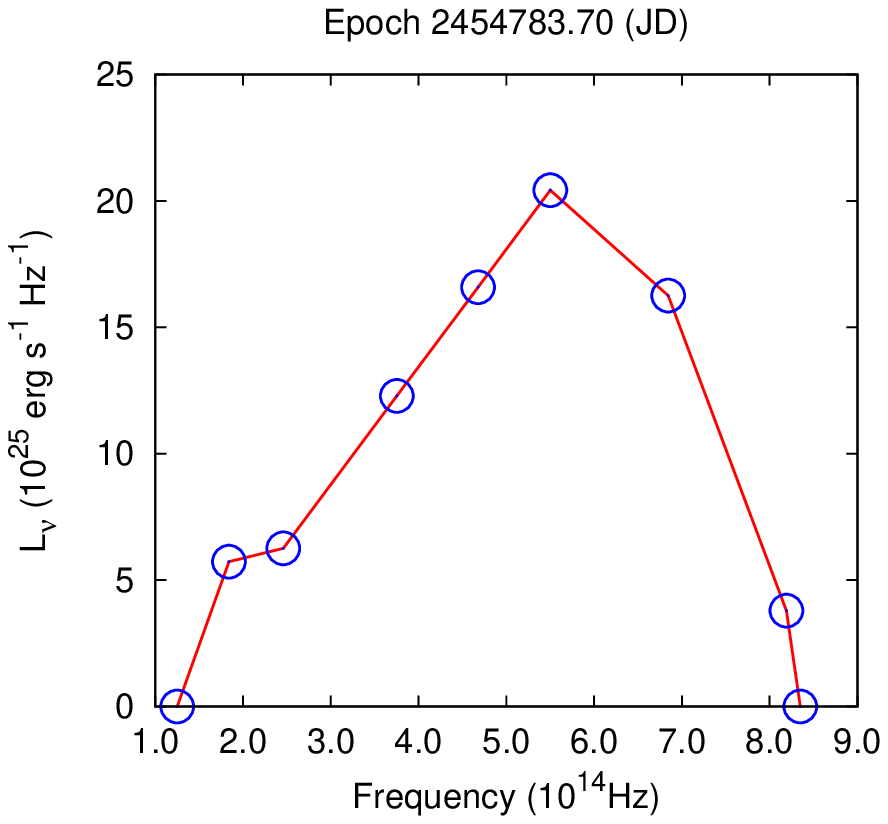}
  \end{minipage}
  \begin{minipage}[t]{0.45\textwidth}
    \epsscale{1.2}
    \plotone{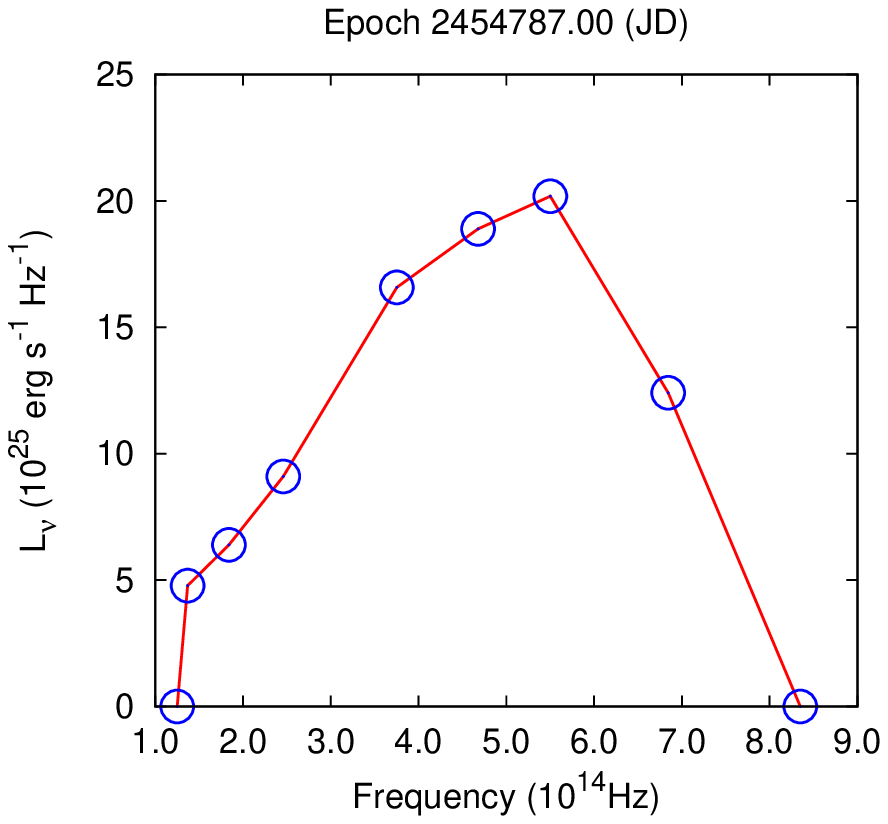}
  \end{minipage}
  \caption{Samples of the SED to construct the bolometric LC.
The top left SED corresponds to the first point of the bolometric LC.
Linear interpolations are made to fill the gaps in observations. The
details of the interpolation are described in the Appendix.
}\label{SED}
\end{center}
\end{figure*}

\section{Discussion and Conclusions}
We have found a fallback SN model (13CO\_2) whose
bolometric LC and photospheric velocity
are both in good agreement with the observations of SN 2008ha.
The ejecta of the model has very low explosion energies
and ejecta masses:
($E_{\mathrm{kin}}$, $M_\mathrm{ej}$) =
($1.2\times10^{48}\ \mathrm{erg}$, $0.074~M_\odot$).
The explosion models from the progenitors of 25CO, 40CO and 40He 
are not in agreement with the
observations of SN 2008ha because the
photospheric velocities of these models are too high to be compatible 
with the observed line velocities of SN 2008ha (Figure \ref{TtoV}).
One might think that a model with sufficiently small \Ekin can have a low
photospheric velocity.
However, as shown in Figure \ref{EtoM}, smaller \Ekin leads to a larger fallback
and thus smaller $M_\mathrm{ej}$ as $M_\mathrm{ej}\propto E_\mathrm{kin}$.
In this case,
the velocity of the ejecta,
$v\propto\left(E_\mathrm{kin}/M_{\mathrm{ej}}\right)^{1/2}$, will not
become smaller.
Alternatively, the explosion
of 25He requires $E_\mathrm{kin}\simeq10^{47}$ erg to have
a rise time similar to SN 2008ha.

The difference between the successful and the unsuccessful progenitor models
to explain SN 2008ha stems from the difference in the density structure.
As discussed in Section 4.1, the density
structure affects the propagation of the shock wave
and thus the relation between \Ekin and $M_\mathrm{ej}$.
This means that whether
a model can have the appropriate $(E_\mathrm{kin}, M_\mathrm{ej})$ for
SN 2008ha mainly depends on the density
structure of the progenitor and not on the progenitor mass.
Thus, the main-sequence mass of SN 2008ha cannot be constrained only by
hydrodynamical calculations
and 13CO would not be a unique progenitor candidate for SN 2008ha.
Other progenitor models with appropriate density structures
are also expected to reproduce
the observational properties of SN 2008ha.

The fact that faint SNe like SN 2008ha could emerge from the
fallback of massive stars is also related to long
gamma-ray bursts (LGRBs) without accompanied SNe (see also V09).
LGRBs are thought to result from the death of massive stars
and the fact that nearby LGRBs are accompanied by bright SNe
(e.g.,
GRB 030329 and SN 2003dh, Hjorth et al. 2003;
GRB 100316D and SN 2010bh, Chornock et al. 2010)
is one of the evidence for the scenario that relates
LGRBs to the death of massive stars.
However, some LGRBs are not accompanied by SNe
even though they are close enough to be observed
(e.g., GRB 060614 (Gehrels et al. 2006; Fynbo et al. 2006;
Della Valle et al. 2006; Gal-Yam et al. 2006))
and this is one of the challenging problems of the theory of LGRBs.
Several scenarios are proposed, e.g., a neutron star-white dwarf
merger (e.g., King et al. 2007) and a massive stellar death
with a faint/dark SN (e.g., Tominaga et al. 2007).
In this paper, we show that core-collapse SNe with fallback
can be very faint and reproduce the observations of a faint
SN (SN 2008ha).
This supports the scenario that
LGRBs without observed bright SNe
are accompanied by very faint SNe with fallback.
In fact, theories of LGRBs like the collapsar model (Woosley 1993)
assume black hole formation with an accretion disk and
this picture is consistent with our fallback SN model in the sense that
most part of progenitors accretes to the central remnant
and the central remnant becomes
massive enough to be a black hole.
This process could induce an LGRB, and
it is possible that an LGRB was actually associated with SN 2008ha.
We could have missed the LGRB because the jet of the LGRB was not
directed to the Earth.
In addition, our fallback SN model for SN 2008ha
requires some mixing process to provide the ejecta with $^{56}$Ni
and the jet from the LGRB could have played a role in the mixing.

\begin{figure*}[t]
\begin{center}
  \begin{minipage}[t]{0.45\textwidth}
    \epsscale{1.2}
    \plotone{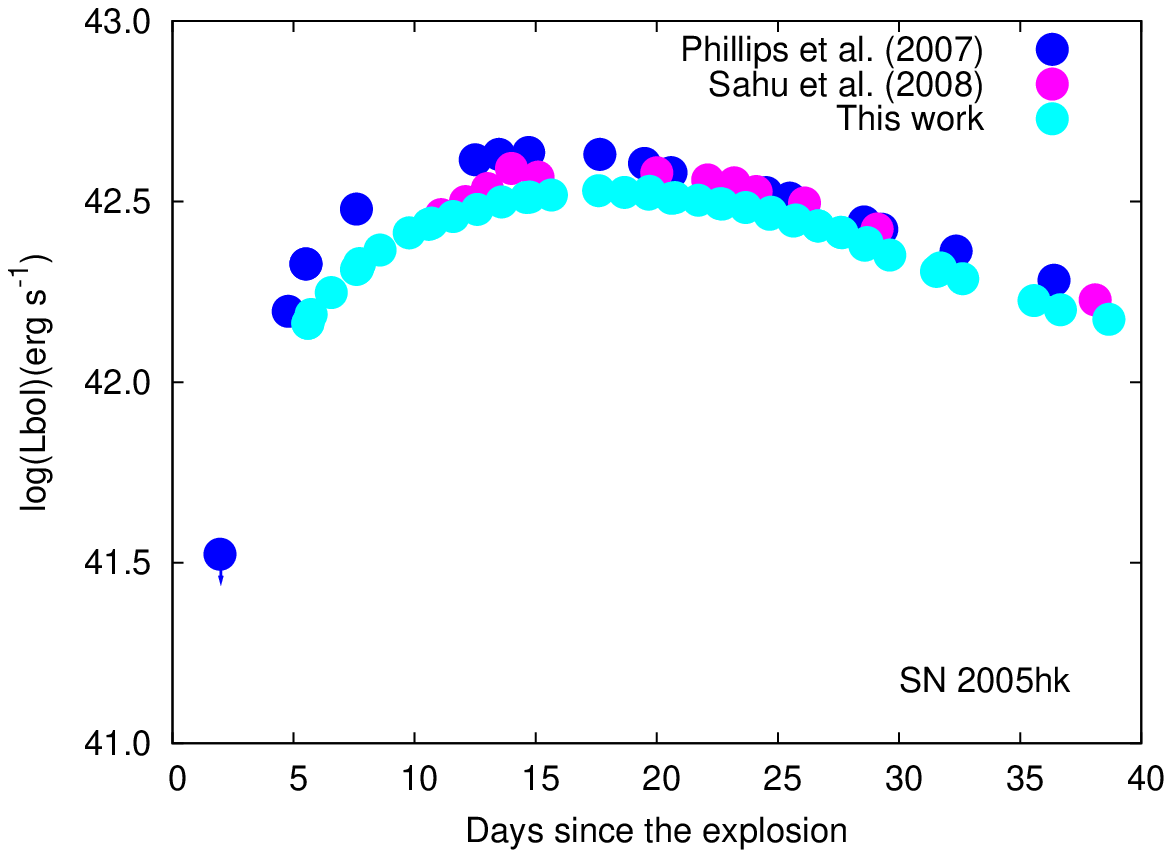}
  \end{minipage}
  \begin{minipage}[t]{0.45\textwidth}
    \epsscale{1.2}
    \plotone{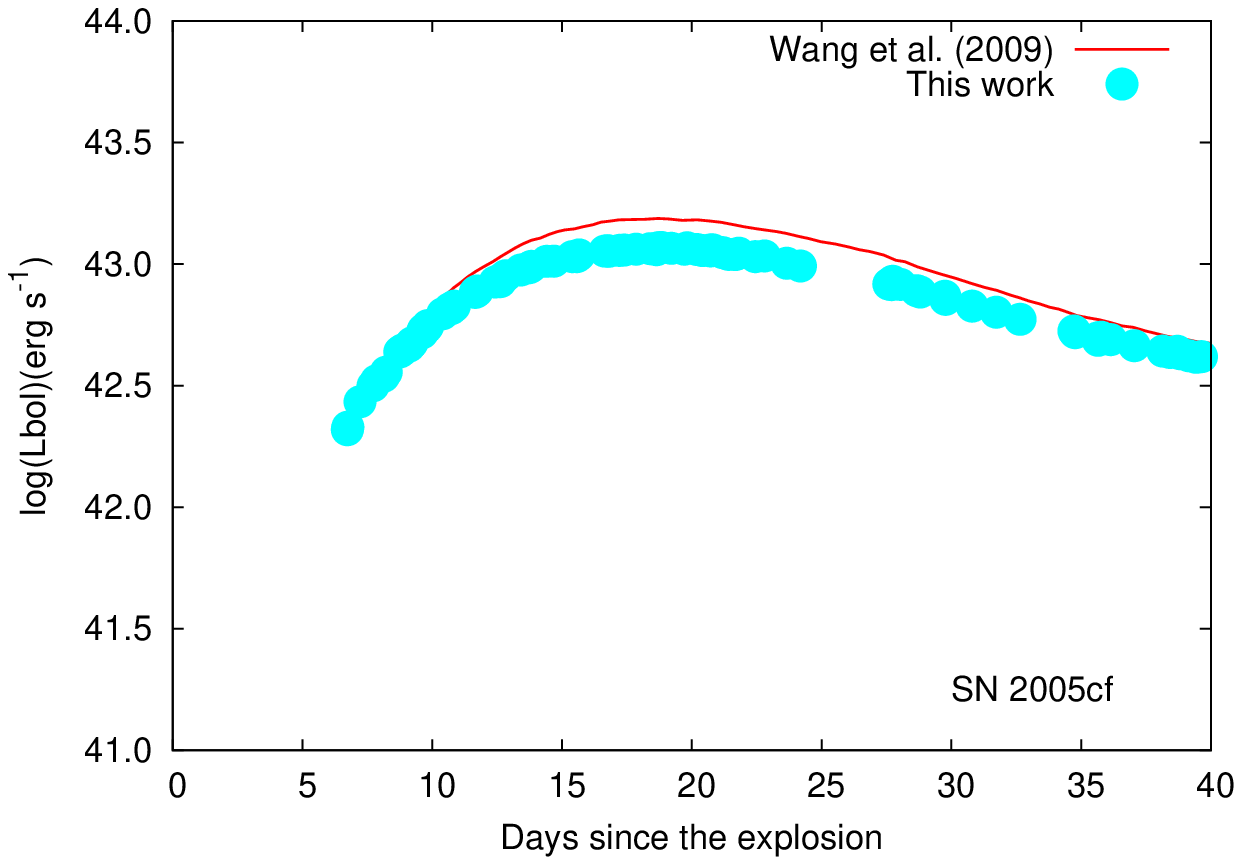}
  \end{minipage}
\caption{
Comparison of the bolometric LCs with those obtained in previous works.
We apply our method to another SN 2002cx-like supernova SN 2005hk (Phillips et
al. 2007, Sahu et al. 2008) and a
normal Type Ia supernova SN 2005cf (Wang et al. 2009).
The rise times of SN 2005hk and SN 2005cf are assumed to be 15 days
 (Phillips et al. 2007) and 18 days (Wang et al. 2009), respectively.
Our bolometric LCs tend to be a bit fainter around the maximum epoch but the
decline rate and the luminosity at later epochs are in good agreement
with other
bolometric LCs obtained in previous works.
}\label{Boltest}
\end{center}
\end{figure*}

\begin{figure}[b]
\begin{center}
\epsscale{1.2}
\rotatebox{0}{
\plotone{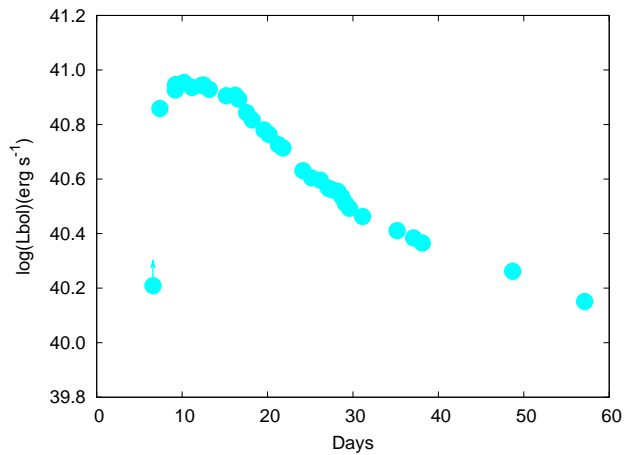}}
\caption{
Obtained bolometric LC of SN 2008ha.
The rise time is set to 10 days in this plot.
}
\label{bol}
\end{center}
\end{figure}

Currently, several SNe are reported to have features similar to SN 2008ha.
SN 2005E is one of the examples (Perets et al. 2010; see also Figure 7 of F09).
As SN 2005E is found far from the disk of the host galaxy,
SN 2005E is unlikely to be a core-collapse SN. However, its spectra
do not show the characteristics of thermonuclear explosions.
Thus, SN 2005E is suggested to be a new type of stellar explosion
(Perets et al. 2010).
Perets et al. (2010) showed that the late phase spectra of SN 2005E
contain strong emission lines of [\ion{Ca}{ii}] $\lambda\lambda$7291, 7323
and suggested that those are the results of 
Ca enrichment in the ejecta. As this feature is
similar to SN 2008ha (Figure 7 of F09), they suggested that
SN 2005E is related to SN 2008ha.
However, the line velocities of SN 2008ha ($\sim 2000\ \kms$)
are considerably slower than SN 2005E ($\sim 11,000\ \kms$).
Thus, it is not so obvious that SN 2005E has the same origin as 
SN 2008ha.
Kawabata et al. (2010) reported 
that Type Ib SN 2005cz is
another example of an SN that shows
prominent Ca emissions and they related it
to the core-collapse SN from a low mass ($\simeq 10~M_\odot$) star.
In this paper, we suggest that SN 2008ha is of core-collapse origin.
Other SN 2002cx-like SNe, SN 2002cx (Li et al. 2003; Jha et al. 2006) and SN
2005hk (Phillips et al. 2007; Sahu et al. 2008), are considered to be weak
thermonuclear explosions while SN 2005E seems to be neither a core-collapse
nor a typical thermonuclear explosion (Perets et al. 2010).
Thus, SN 2002cx-like SNe might contain SNe with
several kinds of origins and other criteria to classify them might be required.
We need more samples of SN 2002cx-like SNe to clarify such criteria.
See the supplementary information of
Kawabata et al. (2010) for more intensive discussion for these Ca-rich
SNe.

We propose that SN 2008ha can 
be a core-collapse SN with fallback.
However, it is not obvious that the features of the intermediate-mass elements
that appeared in the earliest spectrum
of SN 2008ha (Foley et al. 2010) can be synthesized by 
our model and this will be
investigated in a forthcoming paper (D. N. Sauer et al. 2010, in preparation).
Our model does not exclude the possibility that
SN 2008ha is a thermonuclear explosion.
However, simple estimates of \Mej by V09 and F09 are
far below the Chandrasekhar mass limit ($1.4\ M_\odot$),
which is required to ignite a thermonuclear explosion of a white dwarf.
However, these estimates assume that the total opacity $\kappa$ is constant.
It is possible that the effects of the opacity are large enough 
to create the appearance of a thermonuclear explosion similar to
SN 2008ha. If this is the case, the opacity must become
very low to match the fast rise of SN 2008ha as expected from the
relation
$\tau\propto \kappa^{1/2}(M_{\mathrm{ej}}^3/E_{\mathrm{kin}})^{1/4}$.
Other kinds of explosions such as '.Ia' SNe (e.g., Shen et al. 2010)
or
accretion-induced collapses (e.g., Darbha et al. 2010)
might also be candidates but there still do not exist sufficient models
that are consistent with SN 2008ha.

\begin{acknowledgments} 
We thank Sergei I. Blinnikov for useful discussions and comments.
Numerical calculations and data analysis were in part carried out on the general-purpose PC farm at Center for Computational Astrophysics, CfCA, of National Astronomical Observatory of Japan.
This research has been supported in part by World Premier
International Research Center Initiative, MEXT, and by the
Grant-in-Aid for Scientific Research of the JSPS (20540226, 20840007, 21840055)
and MEXT (19047004, 22012003), Japan.
\end{acknowledgments}

\appendix
\section{Construction of Bolometric Light Curve}
We construct the bolometric LC of SN 2008ha by using the $UBVRIJHK$
band LCs
presented by F09. For each
epoch, we derive the spectral energy distribution (SED) and obtain the
bolometric luminosity by integrating the SED.
To estimate the absolute magnitude of each band,
we use the distance modulus $\mu=31.64$ mag and the color excess $E(B-V)=0.08$
mag (F09). Reddening is corrected by
following Cardelli et al. (1989).
For each epoch, we use the observational data if they are available.
If there is no observation of a band at a certain epoch, we deduce
the magnitude of the band
by linearly interpolating the observed magnitude of the nearest epochs.
If no observational data are available before or after the observation
as in the first few epochs when no $RIK$ band data are available,
we linearly interpolate data
points on the SED. Finally, we integrate the SED by
constructing trapezoids and triangles by connecting the $UBVRIJHK$ data
and the two edges.
The two edges are chosen to be $1.25\times10^{14}$ Hz and
$8.35\times10^{14}$ Hz, close to the frequencies of the $K$ band and $U$
band. Some samples of the SED are shown in Figure \ref{SED}.
We calculate the bolometric LCs of SN 2005hk
(Phillips et al. 2007, Sahu et al. 2008) and
SN 2005cf (Wang et al. 2009) to check
our method (Figure \ref{Boltest}).
Our bolometric LCs are in a good agreement with the
bolometric LCs obtained in the previous studies.

The bolometric LC of SN 2008ha obtained by our method is shown in
Figure \ref{bol}.
As the first point is constructed by only the $JH$ data, we assume that the actual
luminosity at this epoch is higher.
The rise time of our bolometric LC is consistent with that of the quasi-bolometric LC constructed by F09.

\bibliographystyle{apj}

\begin{thebibliography}{108}
\expandafter\ifx\csname natexlab\endcsname\relax\def\natexlab#1{#1}\fi

\bibitem[]{}
Arnett, W. D. 1982, \apj, 253, 785

\bibitem[]{}
Aufderheide, M. B., Baron, E., \& Thielemann, T.-K. 1991, \apj, 370, 630

\bibitem[]{}
Axelrod, T. S. 1980, Ph.D. thesis, Univ. California, Santa Cruz

\bibitem[]{}
Branch, D., Baron, E., Thomas, R. C., Kasen, D., Li, W., \& Filippenko,
A. V. 2004, PASP, 116, 903

\bibitem[Bruenn et al.(2009)]{2009AIPC.1111..593B} Bruenn, S.~W., 
Mezzacappa, A., Hix, W.~R., Blondin, J.~M., Marronetti, P., Messer, 
O.~E.~B., Dirk, C.~J., 
\& Yoshida, S.\ 2009, American Institute of Physics Conference Series, 1111, 593

\bibitem[Burrows et al.(2007)]{2007PhR...442...23B} Burrows, A., Dessart, 
L., Ott, C.~D., \& Livne, E.\ 2007, \physrep, 442, 23 

\bibitem[]{}
Cardelli, J. A., Clayton, G. C., \& Mathis, J. S. 1989, \apj, 345, 245

\bibitem[]{}
Chevalier, R. A. 1989, \apj, 346, 847

\bibitem[]{}
Chornock, R., et al. 2010, arXiv:1004.2262

\bibitem[]{}
Colella, P., \& Woodward, P. R. 1984, J. Comput. Phy., 54, 174

\bibitem[]{}
Colgate, S. A. 1971, \apj, 163, 221

\bibitem[]{}
Darbha, S., et al. 2010, arXiv:1005.1081

\bibitem[]{}
Della Valle, M., et al. 2006, Nature, 444, 1050

\bibitem[Foley et al.(2010)]{2010ApJ...708L..61F} Foley, R.~J., Brown, 
P.~J., Rest, A., Challis, P.~J., Kirshner, R.~P., 
\& Wood-Vasey, W.~M.\ 2010, \apjl, 708, L61 

\bibitem[]{}
Foley, R. J., et al. 2009, AJ, 138, 376 (F09)

\bibitem[]{}
Fryer, C. L. 1999, \apj, 522, 413

\bibitem[]{}
Fryer, C. L., Hungerford, A. L., \& Young, P. A. 2007, \apj, 662, L55

\bibitem[Fryer et al.(2009)]{2009ApJ...707..193F} Fryer, C.~L., et al.\ 
2009, \apj, 707, 193 

\bibitem[Fynbo et al.(2006)]{2006Natur.444.1047F} Fynbo, J.~P.~U., et al.\ 
2006, \nat, 444, 1047 

\bibitem[Gal-Yam et al.(2006)]{2006Natur.444.1053G} Gal-Yam, A., et al.\ 
2006, \nat, 444, 1053

\bibitem[Gehrels et al.(2006)]{2006Natur.444.1044G} Gehrels, N., et al.\ 
2006, \nat, 444, 1044 

\bibitem[]{}
Hachisu, I., et al. 1991, \apj, 368, 27 

\bibitem[]{}
Hjorth, J., et al. 2003, Nature, 423, 847

\bibitem[]{}
Iwamoto, N., Umeda, H., Tominaga, N., Nomoto, K., \& Maeda, K. 2005, Science, 309, 451 

\bibitem[]{}
Iwamoto, K., et al. 2000, \apj, 534, 660 

\bibitem[]{}
Janka, H.-Th., Langanke, K., Marek, A., Mart$\rm{\acute{\i}}$nez-Pinedo, G., \&
M$\rm{\ddot{u}}$ller, B. 2007, Phys. Rep., 442, 38

\bibitem[]{}
Jha, S., et al. 2006, AJ, 132, 189

\bibitem[Joggerst et al.(2009)]{2009ApJ...693.1780J} Joggerst, C.~C., 
Woosley, S.~E., \& Heger, A.\ 2009, \apj, 693, 1780 

\bibitem[Kawabata et al.(2009)]{2009arXiv0906.2811K} Kawabata, K.~S., 
et al. 2010, Nature, 465, 326

\bibitem[King et al.(2007)]{2007MNRAS.374L..34K} King, A., Olsson, E., 
\& Davies, M.~B.\ 2007, \mnras, 374, L34 

\bibitem[]{}
Li, W., et al. 2003, PASP, 115, 453

\bibitem[]{}
MacFadyen, A. I., Woosley, S. E., \& Heger, A. 2001, \apj, 550, 410

\bibitem[]{}
Maeda, K., Mazzali, P. A., Deng, J., Nomoto, K., Yoshii, Y., Tomita, H., \&
Kobayashi, Y. 2003a, \apj, 593, 931

\bibitem[]{}
Maeda, K., \& Nomoto, K. 2003b, ApJ, 598, 1163

\bibitem[]{}
Nadezhin, D. K., \& Frank-Kamenetskii, D. A. 1963, Soviet Astro.-AJ, 6, 779

\bibitem[]{}
Nakamura, T., et al. 2001, \apj, 555, 880

\bibitem[]{}
Nomoto, K. 1982, \apj, 253, 798

\bibitem[Nomoto(1984)]{1984ApJ...277..791N} Nomoto, K.\ 1984, \apj, 277, 
791 

\bibitem[]{}
Nomoto, K., \& Hashimoto, M. 1988, Phys. Rep., 163, 13 

\bibitem[]{}
Perets, H. B., et al. 2010, Nature, 465, 322

\bibitem[]{}
Phillips, M. M., et al. 2007, PASP, 119, 360

\bibitem[]{}
Puckett, T., Moore, C., Newton, J., \& Orff, T. 2008,
Central Bureau Electronic Telegrams, 1567, 1

\bibitem[Pumo et al.(2009)]{2009ApJ...705L.138P} Pumo, M.~L., et al.\ 2009, 
\apjl, 705, L138 

\bibitem[]{}
Sahu, D. K., Tanaka, M., Anupama, G. C., Kawabata, K. S., Maeda, K.,
Tominaga, N., Nomoto, K., Mazzali, P. A., \& Prabhu, T. P. 2008,
\apj, 680, 580 

\bibitem[]{}
Sedov, L. 1959, Similarity and Dimensional Methods in Mechanics
(New York: Academic)

\bibitem[]{}
Shen, K. J., Kasen, D., Weinberg, N. N., Bildsten, L.,
\& Scannapieco, E. 2010, \apj, 715, 767

\bibitem[]{}
Sugimoto, D., \& Nomoto, K. 1975,
Sci. Pap. Coll. Gen. Duc. Univ. Tokyo, 25, 109

\bibitem[Suwa et al.(2009)]{2009arXiv0912.1157S} Suwa, Y., Kotake, K., 
Takiwaki, T., Whitehouse, S.~C., Liebendoerfer, M., 
\& Sato, K.\ 2009, arXiv:0912.1157 

\bibitem[]{}
Tominaga, N. 2009, \apj, 690, 526

\bibitem[Tominaga et al.(2007)]{2007ApJ...657L..77T} Tominaga, N., Maeda, 
K., Umeda, H., Nomoto, K., Tanaka, M., Iwamoto, N., Suzuki, T., 
\& Mazzali, P.~A.\ 2007, \apjl, 657, L77 

\bibitem[]{}
Umeda, H., \& Nomoto, K. 2002, \apj, 565, 385

\bibitem[]{}
Umeda, H., \& Nomoto, K. 2005, \apj, 619, 427

\bibitem[]{}
Valenti, S., et al. 2009, Nature, 459, 674 (V09)

\bibitem[]{}
Wang, X., et al. 2009, \apj, 697, 380

\bibitem[]{}
Woosley, S. E. 1993, \apj, 405, 273

\bibitem[]{}
Woosley, S. E., \& Weaver, T. A. 1995, ApJS, 101, 181

\bibitem[]{}
Young, A. P., \& Fryer, C. L. 2007, ApJ, 664, 1033

\bibitem[]{}
Zhang, W., Woosley, S. E., \& Heger, A. 2008, \apj, 679, 639

\end{thebibliography}

\end{document}